\documentclass[aoas,MSNbibl,nameyear,dvips]{arximspdf}
\usepackage{dcolumn,multirow}
\usepackage{graphicx}

\doi{10.1214/14-AOAS773} 
\volume{8}
\issue{4}
\pubyear{2014}
\firstpage{2336}
\lastpage{2355}
\docsubty{FLA}
\setattribute{copyright}{owner}{In the Public Domain}

\makeatletter
\newcommand{\rrVert}{\Vert}
\newcommand{\llVert}{\Vert}
\newcolumntype{d}[1]{D{.}{.}{#1}}
\newcommand{\eqref}[1]{(\ref{#1})}
\newcommand{\sd}{\operatorname{sd}}
\newcommand{\pr}{\mathrm{P}}
\newcommand{\epn}{\mathrm{E}}
\newcommand{\bmX}{\mathbf X}
\newcommand{\bmW}{\mathbf W}
\newcommand{\bmalpha}{\bolds{\alpha}}
\newcommand{\bmbeta}{\bolds{\beta}}
\newcommand{\bmtheta}{\bolds{\theta}}
\makeatother

\begin{document}
\begin{frontmatter}

\title{The use of covariates and random effects in evaluating
predictive biomarkers under a~potential~outcome framework}
\runtitle{Evaluating predictive biomarkers}

\begin{aug}
\author[A]{\fnms{Zhiwei}~\snm{Zhang}\corref{}\ead[label=e1]{zhiwei.zhang@fda.hhs.gov}\thanksref{M1}},
\author[B]{\fnms{Lei}~\snm{Nie}\thanksref{M2}\ead[label=e2]{lei.nie@fda.hhs.gov}},
\author[C]{\fnms{Guoxing}~\snm{Soon}\thanksref{M2}\ead[label=e3]{guoxing.soon@fda.hhs.gov}}
\and
\author[D]{\fnms{Aiyi}~\snm{Liu}\thanksref{M3}\ead[label=e4]{liua@mail.nih.gov}}
\runauthor{Zhang, Nie, Soon and Liu}
\affiliation{Center for Devices and Radiological Health, Food and Drug Administration\thanksmark{M1},
Center for Drug Evaluation and Research, Food and Drug Administration\thanksmark{M2}
and
\textit{Eunice Kennedy Shriver} National Institute of Child Health
and Human Development\thanksmark{M3}}
\address[A]{Z. Zhang\\
Division of Biostatistics\\
Office of Surveillance and Biometrics\\
Center for Devices\\
\quad and Radiological Health,\\
\quad Food and Drug Administration\\
10903 New Hampshire Ave\\
Silver Spring, Maryland 20993\\
USA\\
\printead{e1}}
\address[B]{L. Nie\\
Division of Biometrics V\\
Office of Biostatistics\\
Center for Drug Evaluation and Research,\hspace*{1pt}\\
\quad Food and Drug Administration\\
10903 New Hampshire Ave\\
Silver Spring, Maryland 20993\\
USA\\
\printead{e2}}
\address[C]{G. Soon\\
Division of Biometrics IV\\
Office of Biostatistics\\
Center for Drug Evaluation and Research,\\
\quad Food and Drug Administration\\
10903 New Hampshire Ave\\
Silver Spring, Maryland 20993\\
USA\\
\printead{e3}}
\address[D]{A. Liu\\
Biostatistics and Bioinformatics Branch\\
Division of Intramural Population\\
\quad Health Research\\
{Eunice Kennedy Shriver} National Institute\\
\quad of Child Health and Human Development\\
National Institutes of Health\\
6100 Executive Blvd.\\
Bethesda, Maryland 20892\\
USA\\
\printead{e4}}
\end{aug}

\received{\smonth{1} \syear{2013}}
\revised{\smonth{4} \syear{2014}}

%
\begin{abstract}
Predictive or treatment selection biomarkers are usually evaluated in a~subgroup 
or regression analysis with focus on the treatment-by-marker
interaction. Under a potential outcome framework (Huang, Gilbert and
Janes [\textit{Biometrics} \textbf{68} (2012) 687--696]), a predictive\break 
biomarker is considered a predictor for a desirable treatment benefit
(defined by comparing potential outcomes for different treatments) and
evaluated using familiar concepts in \mbox{prediction} and classification.
However, the desired treatment benefit is unobservable because each
patient can receive only one treatment in a typical study. Huang et
al.~overcome this problem by assuming monotonicity of potential
outcomes, with one treatment dominating the other in all patients.
\mbox{Motivated} by an HIV example that appears to violate the monotonicity
assumption, we propose a different approach based on covariates and
random effects for evaluating predictive biomarkers under the potential
outcome framework. Under the proposed approach, the parameters of
interest can be identified by assuming conditional independence of
potential outcomes given observed covariates, and a sensitivity
analysis can be performed by incorporating an unobserved random effect
that accounts for any residual dependence. Application of this approach
to the motivating example shows that baseline viral load and CD4 cell
count are both useful as predictive biomarkers for choosing
antiretroviral drugs for treatment-naive patients.
\end{abstract}

%
\end{frontmatter}

\section{Introduction}\label{intro}

Much of contemporary medical research is focused on treatment effect
heterogeneity, that is, the fact that the same treatment can have
different effects on different patients. The increasing awareness of
treatment effect heterogeneity has motivated the development of
predictive biomarkers for identifying the subpopulation of patients who
would actually benefit from a new treatment [e.g., \citeauthor{s08} (\citeyear{s08,s10})]. 
Classical examples of predictive biomarkers include genetic
markers for cancer treatment, such as the OncoType Dx multi-gene score
for breast cancer [\citet{p04}] and the K-RAS gene expression
level for colorectal cancer [\citet{k08}]. This article is
motivated by a new and growing interest in the possibility of using
baseline viral load or CD4 cell count as a predictive biomarker for
treating human immunodeficiency virus type 1 \mbox{(HIV-1)}.

%
\begin{figure}

\includegraphics{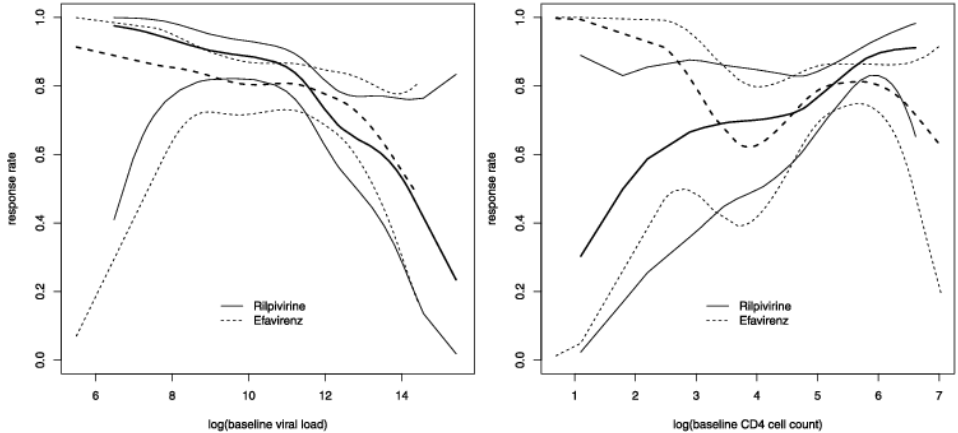}

\caption{Nonparametric regression analysis of the THRIVE data: smoothed
estimates (thicker lines) and pointwise 95\% confidence intervals
(thinner lines) for treatment-specific response rates as functions of
baseline viral load (left) and CD4 cell count (right).}\label{fig1}
\end{figure}

Evaluation of a predictive biomarker is usually based on a subgroup or
regression analysis comparing treatment effects on different
subpopulations defined by the biomarker [e.g., \citet{g85,r97,p02}]. Under this approach, the performance of a predictive
biomarker is measured by the interaction between marker value and
treatment assignment in a regression model for the clinical outcome of
interest, which will be referred to as an outcome model. For example,
consider the THRIVE study, a phase 3, randomized, noninferiority trial
comparing rilpivirine, a newly developed nonnucleoside reverse
transcriptase inhibitor, with efavirenz in treatment-naive adults
infected with HIV-1 [\citet{c11}]. The outcome of primary interest
is a binary indicator of virologic response at week 48 of treatment
(see Section~\ref{app} for details). Our consideration of baseline
viral load and CD4 cell count as predictive biomarkers (for choosing
between efavirenz and rilpivirine) is motivated by Figure~\ref{fig1}, which
shows nonparametric estimates of treatment-specific response rates as
functions of marker value, separately for each biomarker. Figure~\ref{fig1}
suggests a qualitative interaction between treatment and each
biomarker, with rilpivirine favored over efavirenz for higher values of
baseline CD4 cell count and lower values of baseline viral load.
Because the statistical significance in Figure~\ref{fig1} is not straightforward
to assess, a simple logistic regression analysis that includes
treatment, marker and their interaction is performed (separately for
each marker), and the resulting $p$-value for the treatment-by-marker
interaction is 0.059 for viral load and 0.031 for CD4 cell count.

While informative about possible interactions, Figure~\ref{fig1} is less
transparent about the predictive performance of these biomarkers and
their comparison. \citet{h12} point out that a strong interaction
is not sufficient for adequate performance of a predictive biomarker,
that the scale of the interaction coefficient depends on the functional
form of the outcome model, and that the interaction-based approach is
ill-suited for developing combination markers. These authors also
propose a potential outcome framework where a predictive biomarker
is---as the term suggests---treated as a predictor for a desirable
treatment benefit. Note that a treatment benefit is necessarily the
result of comparing potential outcomes for different treatments applied
to the same patient. Under this perspective, predictive biomarkers
should be evaluated using familiar measures in prediction and
classification [e.g., \citet{z02,p03,z11}]. Specifically, one
should consider the true and false positive rates of a binary marker
and the receiver operating characteristic (ROC) curve for a continuous
marker. This approach allows different markers to be compared on the
same scale and facilitates the development of combination markers.

The objective of this article is to evaluate and compare baseline viral
load and CD4 cell count as predictive biomarkers under the potential
outcome framework. It is important to distinguish this objective from
the related problems of identifying potential markers, combining
several markers into a hybrid marker, choosing the cutoff point for a
given marker and, more generally, developing an individualized
treatment strategy. Variable selection techniques such as lasso-based
methods have been used to select and combine genetic markers [e.g.,
\citet{t12}]. Nonparametric multivariate methods [e.g.,
\citet
{su08,f11,q11}] and semiparametric methods [e.g., \citet{z12}]
have been used to develop a treatment rule (i.e., a set of criteria for
selecting patients), which can be considered a binary hybrid marker
obtained by combining multiple markers. A natural question that arises
from the THRIVE study is whether or how to combine baseline viral load
and CD4 cell count into a hybrid marker with improved predictive
accuracy. While that question is beyond the scope of this article, we
note that the potential outcome framework and the proposed methods are
applicable to any given marker. Once a hybrid marker is developed, it
can be evaluated and compared to the individual markers in the same
framework using the same methods.

An analytical challenge for the potential outcome framework is that the
desired treatment benefit, which involves potential outcomes under
different treatments, is usually unobservable because each patient can
receive only one treatment in a typical study. A possible exception to
this limitation is a cross-over study, which has its own issues [e.g.,
\citet{p12}] and will not be discussed in this article. In a
typical clinical study such as the THRIVE study, the standard
methodology in prediction and classification is not directly applicable
to a predictive biomarker. To address this issue and the resulting
identification problem, \citet{h12} make a monotonicity
assumption, namely, that one treatment dominates the other in all
individual patients, and suggest a sensitivity analysis for possible
departures from the monotonicity assumption. While the monotonicity
assumption may be plausible in some situations such as vaccine trials,
it may be less appealing as a starting point in other situations. In
the THRIVE study, for example, the presence of a qualitative
interaction (in the sense that neither treatment has a higher response
rate for all realistic marker values) implies that neither treatment is
dominant in all patients. Additionally, the approach of \citet
{h12} is developed for a binary outcome and not readily extensible to
other types of outcomes.

In this article, we propose alternative methods that do not require
monotonicity or assume a binary outcome. Our first step is to account
for the dependence between potential outcomes (for different treatments
applied to the same patient) by adjusting for relevant covariates, such
as demographic variables and baseline characteristics. If the set of
measured covariates is sufficient for explaining the dependence between
potential outcomes, the aforementioned performance measures can then be
identified by assuming conditional independence of potential outcomes
given covariates. Possible violations of this assumption can be
addressed by introducing a random effect to account for residual
dependence. In the next section we formulate the problem in terms of
potential outcomes and provide a general rationale for the proposed
approach. The proposed methods are then described in Section~\ref{meth}
and applied to the THRIVE study in Section~\ref{app}. The article ends
with a discussion in Section~\ref{disc}. Some technical details are
provided in a supplemental article [\citet{z14}].

\section{Notation and rationale}\label{nr}

Suppose a randomized clinical trial is conducted to compare an
experimental treatment with a control treatment, which may be a placebo
or a standard treatment, with respect to a clinical outcome of
interest, which may be discrete or continuous. Because there is only
one outcome of primary interest in the THRIVE study, we will work with
a scalar outcome unless otherwise noted. However, most of our
methodology is readily applicable to a vector-valued outcome, with the
exception of the sensitivity analysis in Section~\ref{indir} (described
fully in Web Appendix B). For a generic patient in the target
population, let $Y(t)$ denote the potential outcome that will result if
the patient receives treatment $t$ (0 for control; 1 for experimental).
Note that the $Y(t)$, $t=0,1$, cannot both be observed at a given
time. Let $T$ denote the treatment assigned randomly to a study
subject, thus $T$ is a Bernoulli variable independent of all baseline
variables. Without considering noncompliance, which is negligible in
the THRIVE study, we assume that $T$ is also the actual treatment given
to the subject, and write $Y=Y(T)$ for the actual outcome. Should
noncompliance become a major issue, we could take an intent-to-treat
perspective and compare treatment assignments or use analytical
techniques to recover the actual treatment effect [e.g., \citet
{v03}]. We assume that large values of $Y$ are desirable. Where
necessary, the subscript $i=1,\ldots,n$ will be attached to random
variables to denote individual patients in the trial.

Our interest is in evaluating a predictive biomarker $Z$, a baseline
variable which may be binary or continuous, with higher marker values
supporting the use of the new treatment. The biomarker $Z$ is intended
to identify the subpopulation of patients who would benefit from the
new treatment relative to the control. It can be a continuous variable
as in our motivating example or a binary one such as a treatment rule
developed using nonparametric multivariate methods. Let the desired
treatment benefit be indicated by $B=I\{(Y(0),Y(1))\in\mathcal B\}$,
where $I\{\cdot\}$ is the indicator function and $\mathcal B$ is the
set of desirable outcomes. Note that $B$ is by definition a comparison
of the two potential outcomes. For a binary outcome, $B$ might be an
indicator for $Y(0)<Y(1)$ or $Y(0)\le Y(1)$, depending on which
treatment is preferable with identical (efficacy) outcomes. In our
example, we set $B=I\{Y(0)\le Y(1)\}$ because rilpivirine is thought to
have a better safety profile and will likely be preferred over
efavirenz when their efficacy outcomes are identical. For a continuous
outcome, we might take $\mathcal B=\{(y_0,y_1)\dvtx y_1-y_0>\delta\}$, where
$\delta$ reflects considerations of cost, clinical significance and
possibly the safety profiles of the two treatments (if not incorporated
into a vector-valued outcome). For an ordered categorical outcome, the
definition of $B$ may be more complicated. We shall take the definition
of $B$ as given and focus on the evaluation of $Z$ for predicting $B$.
The target $B$ is an intrinsic characteristic of an individual patient,
which suggests that $Z$ can be evaluated using well-known quantities in
prediction and classification [e.g., \citet{z02,p03,z11}]. For a
binary marker, it makes sense to consider the true and false positive
rates, defined as $\mathrm{TPR}=\pr(Z=1|B=1)$ and $\mathrm{FPR}=\pr
(Z=1|B=0)$, respectively. For a continuous marker, it is customary to
consider the ROC curve defined as
\[
\label{20} \mathrm{ROC}(s)=1-F_{Z|B=1}\bigl\{F_{Z|B=0}^{-1}(1-s)
\bigr\}.
\]
Here and in the sequel, we use $F$ to denote a generic (conditional)
distribution function, with the subscript indicating the random
variable(s) concerned. The ROC curve is simply a plot of TPR versus FPR
for classifiers of the form $I(Z>z)$, with the threshold $z$ ranging
over all possible values.

Because $B$ is never observed, the existing methodology for evaluating
predictors, which generally assumes that $B$ can be observed, cannot be
used directly to evaluate a predictive biomarker. Nonetheless, we note
that TPR, FPR and ROC are all determined by $F_{Z|B}$, which can be
recovered using Bayes' rule from the marginal distribution of $Z$ and
the conditional probability $\pi_Z(z)=\pr(B=1|Z=z)$. For a binary
marker, $1-\pi_Z(0)$ and $\pi_Z(1)$ are negative and positive
predictive values, respectively,
%
%
\begin{eqnarray}
\label{30} \mathrm{TPR}&=&\frac{\tau\pi_Z(1)}{\tau\pi
_Z(1)+(1-\tau)\pi_Z(0)},
\nonumber\\[-8pt]\\[-8pt]\nonumber
\mathrm{FPR}&=&\frac{\tau\{1-\pi_Z(1)\}}{\tau\{1-\pi_Z(1)\}
+(1-\tau)\{
1-\pi_Z(0)\}},
\end{eqnarray}
where $\tau=\pr(Z=1)$. For a continuous marker, we have
%
%
\begin{eqnarray}
\label{40} F_{Z|B=1}(z_0)&=&\int_{-\infty}^{z_0}
\pi_Z(z)\,dF_Z(z) \Big/\int_{-\infty}^{\infty}
\pi_Z(z)\,dF_Z(z),
\nonumber\\[-8pt]\\[-8pt]\nonumber
F_{Z|B=0}(z_0)&=&\int_{-\infty}^{z_0}
\bigl\{1-\pi_Z(z)\bigr\}\,dF_Z(z) \Big/\int
_{-\infty}^{\infty}\bigl\{1-\pi_Z(z)\bigr
\}\,dF_Z(z).
\end{eqnarray}
Since $Z$ is fully observed, the identifiability of $F_{Z|B}$ would
follow from that of $\pi_Z(z)$. Once an estimate of $\pi_Z(z)$ is
available, it can be substituted into the above displays together with
an empirical estimate of $\tau$ or $F_Z$, depending on the nature of $Z$.

Despite its simple appearance, $\pi_Z(z)$ is not straightforward to
estimate. In fact, for any conceivable form of $\mathcal B$, the
quantity $\pi_Z(z)=\pr\{(Y(0),Y(1))\in\mathcal B|Z=z\}$ is not
empirically identifiable because it involves the joint distribution of
$Y(0)$ and $Y(1)$ given $Z=z$. Owing to randomization, it is
straightforward to identify
\[
F_{t|Z}(y|z)=\pr\bigl\{Y(t)\le y|Z=z\bigr\}=\pr(Y\le y|T=t,Z=z)
\]
for each $t\in\{0,1\}$, and to estimate it from a regression analysis
for $Y$ given $T$~and~$Z$. However, the dependence structure of $Y(0)$
and $Y(1)$ given $Z=z$ is not identifiable from the data [e.g.,
\citet{g00}], which is also known as the fundamental problem of
causal inference [\citet{h86}]. Because $\pi_Z(z)$ is not
determined by the ``marginals'' $F_{t|Z}$ ($t=0,1$), its identification
and estimation require additional information or assumptions about the
dependence between $Y(0)$ and $Y(1)$ given $Z=z$. For a binary outcome,
this can be achieved by assuming monotonicity [i.e., $Y(0)\le Y(1)$
with probability 1] as in \citet{h12}. The monotonicity assumption
corresponds to maximal positive dependence of $Y(0)$ and $Y(1)$.

For a general outcome and without assuming monotonicity, we develop
alternative methods by adapting the techniques of \citet{d03} and
\citet{z13}. To account for the dependence of $Y(0)$ and $Y(1)$,
we start by conditioning on relevant covariates that are associated
with both outcomes. Let $\bmX$ denote a vector of such covariates
measured at baseline, which may include prognostic factors and effect
modifiers. In the THRIVE study, $\bmX$ may include gender, race, and
baseline age and body mass index. We include $Z$ as a component of
$\bmX
$ and write $\bmX=(Z,\bmW)$, where $\bmW$ consists of the additional
baseline covariates. Writing $\pi_X(\mathbf x)=\pr\{
(Y(0),Y(1))\in\mathcal
B|\bmX=\mathbf x\}$, a conditioning argument yields
%
%
\begin{equation}
\label{50} \pi_Z(z)=\epn\bigl\{\pi_X(\bmX)|Z=z\bigr
\}=\int\pi_X(z,\mathbf w)F_{W|Z}(d\mathbf w|z).
\end{equation}
Because $F_{W|Z}$ is empirically identifiable and estimable, the
challenge now is to identify and estimate $\pi_X(\mathbf x)$.

If $\bmX$ is sufficient for explaining the dependence between $Y(0)$
and $Y(1)$, then we can expect that
%
%
\begin{equation}
\label{ci} Y(0)\perp Y(1)|\bmX,
\end{equation}
that is, that $Y(0)$ and $Y(1)$ are conditionally independent given
$\bmX$. This assumption cannot be verified with the observed data and
must be based on external information. Under assumption \eqref{ci}, the
joint distribution of $Y(0)$ and $Y(1)$ given $\bmX$ is determined by
the ``marginals''
\[
F_{t|X}(y|\mathbf x)=\pr\bigl\{Y(t)\le y|\bmX=\mathbf x\bigr
\}=\pr(Y\le y|T=t,
\bmX=\mathbf x)\qquad(t=0,1),
\]
which are straightforward to identify and estimate.
In reality, assumption \eqref{ci} can be violated because $\bmX$ may
not explain all the dependence between $Y(0)$ and $Y(1)$. Such
violations can be examined in a sensitivity analysis based on a latent
variable that accounts for any residual dependence between $Y(0)$ and
$Y(1)$. Under this approach, assumption \eqref{ci} is relaxed as follows:
%
%
\begin{equation}
\label{cire} Y(0)\perp Y(1)|(\bmX,U),
\end{equation}
where $U$ is a subject-specific latent variable that is independent of
$\bmX$. In other words, $U$ represents what is missing from $\bmX$ that
makes assumption \eqref{ci} break down. Assumption \eqref{cire} alone
is not sufficient to identify $\pi_X(\mathbf x)$ because $U$ is unobserved.
However, by specifying certain quantities related to $U$, we can
perform a sensitivity analysis based on assumption \eqref{cire}, as we
demonstrate in the next section.

\section{Methodology}\label{meth}

We now describe methods for estimating the aforementioned performance
measures (TPR, FPR and ROC). As indicated earlier, we will start by
estimating $\pi_X(\mathbf x)$ under assumptions \eqref{ci} and
\eqref{cire}.
This can be done using a direct approach and an indirect approach, to
be described in Sections~\ref{dir} and \ref{indir}, respectively. The
direct and indirect approaches are based on models for $\pi
_X(\mathbf x)$
and $F_{Y|T,X}(y|t,\mathbf x)$, respectively, which we refer to as benefit
and outcome models. A benefit model is directly informative about $\pi
_X(\mathbf x)$ and thus more interpretable in the present context,
while an
outcome model is more familiar to practitioners and easier to estimate
and validate using standard techniques. Further comments comparing the
two approaches are given at the end of Section~\ref{indir} after the
approaches are described and in Section~\ref{disc}. In Section~\ref
{bet} we show how to convert an estimate of $\pi_X(\mathbf x)$
into one of
$\pi_Z(z)$. Estimates of the performance measures of interest are given
in Section~\ref{poi}.

\subsection{Direct estimation of \texorpdfstring{$\pi_X(\mathbf x)$}{piX(x)} based on a Benefit model}\label{dir}

A benefit model is a parametric model for $\pr(B=1|\bmX)=\pr\{
(Y(0),Y(1))\in\mathcal B|\bmX\}$, such as the following generalized
linear model (GLM):
%
%
\begin{equation}
\label{520} \pi_X(\bmX;\bmalpha)=\psi\bigl(\alpha_1+
\bmalpha_X'\bmX\bigr),
\end{equation}
where $\bmalpha=(\alpha_1,\bmalpha_X')'$ is the regression parameter
and $\psi$ is an inverse link function. Since $B$ is binary, the probit
and logit links are natural choices.

Suppose the conditional independence assumption \eqref{ci} holds. To
gain some intuition, consider a discrete $\bmX$ taking values in $\{
\mathbf x
_1,\ldots,\mathbf x_K\}$. Within each stratum defined by $\bmX
=\mathbf x_k$,
assumption \eqref{ci} implies that $Y(0)$ and $Y(1)$ are independent of
each other, as if they arise from different subjects, which we assume
are independent. In other words, given that $\bmX_i=\bmX_j$, the
natural pair $(Y_{i}(0),Y_{i}(1))$ is identically distributed as the
artificial pair $(Y_{i}(0),Y_{j}(1))$. If $T_i=0$ and $T_j=1$, then
$(Y_{i}(0),Y_{j}(1))$ is observable as $(Y_i,Y_j)$, so that $\pi
_X(\mathbf x
_k)=\epn(B|\bmX=\mathbf x_k)$ can be estimated by
%
%
\begin{equation}
\label{510} \frac{1}{n_{0k}n_{1k}}\sum_{i\in\mathcal S_{0k}}\sum
_{j\in\mathcal S_{1k}}B_{ij},
\end{equation}
where $B_{ij}=I\{(Y_i,Y_j)\in\mathcal B\}$, $\mathcal S_{tk}=\{
i\dvtx T_i=t,\bmX_i=\mathbf x_k\}$, and $n_{tk}$ denotes the size of
$\mathcal
S_{tk}$ ($t=0,1$; $k=1,\ldots,K$).

The question is how to generalize this idea to a nondiscrete $\bmX$. To
follow the logic of \eqref{510}, one would need to find subjects from
different treatment groups with the same value of $\bmX$, which becomes
difficult when $\bmX$ has continuous components (4 in our example). To
overcome that problem, we borrow ideas from \citet{d03}, who
consider semiparametric regression for the area under the ROC curve,
and work with an expanded model given by
%
%
\begin{equation}
\label{530} \pr(B_{ij}=1|\bmX_i,\bmX_j;
\bmbeta)=\psi\bigl\{\beta_1+\bolds{\beta}_X'(
\bmX_i+\bmX_j)/2+\bolds{\beta}_{dX}'(
\bmX_j-\bmX_i)\bigr\},
\end{equation}
where $\bmbeta=(\beta_1,\bmbeta_X',\bmbeta_{dX}')'$, $i\in\mathcal
S_{0k}$ and $j\in\mathcal S_{1k}$. The new features of model \eqref
{530} relative to model \eqref{520} are introduced for the sole purpose
of estimating $\bmalpha$. Our research question does not pertain to the
left-hand side of \eqref{530} or the regression coefficient~$\bmbeta
_{dX}$. However, assumption \eqref{ci} implies that $\pr
(B_{ij}=1|\bmX
_i=\bmX_j=\mathbf x)=\pr(B=1|\bmX=\mathbf x)$. Thus, when
$\bmX_i=\bmX_j$, model
\eqref{530} reduces to model \eqref{520} with
%
%
\begin{equation}
\label{535} \bmalpha=\bigl(\beta_1,\bolds{\beta}_X'
\bigr)'.
\end{equation}
In that sense, model \eqref{530} is a helper model that allows us to
estimate model \eqref{520} with the observed data. Let $\mathcal S_t=\{
i\dvtx T_i=t\}$ and let $n_t$ denote the size of $\mathcal S_t$ ($t=0,1$).
Then the regression parameter $\bmbeta$ in model \eqref{530} can be
estimated by solving the equation
%
%
\begin{equation}
\label{540} 0=\sum_{i\in\mathcal S_0}\sum
_{j\in\mathcal S_1}\frac{\partial\pi_{ij}}{\partial\bmbeta}\frac
{B_{ij}-\pi_{ij}}{\pi_{ij}(1-\pi_{ij})},
\end{equation}
where
\[
\pi_{ij}=\psi\bigl\{\beta_1+\bmbeta_X'(\bmX_i+\bmX_j)/2+\bmbeta
_{dX}'(\bmX
_j-\bmX_i)\bigr\}.
\]
As suggested by \citet{d03}, equation \eqref{540} need not include
all possible pairs $(i,j)$; it could be based on a subset of pairs such
that $\llVert\bmX_i-\bmX_j\rrVert<\varepsilon$, where $\llVert
\cdot\rrVert$ denotes Euclidean norm, for some $\varepsilon>0$. The
choice of $\varepsilon$ represents a bias-variance trade-off, where a
larger $\varepsilon$ leads to better efficiency and stability and also
more sensitivity to the last component of model \eqref{530}.

The approach just described relies heavily on the conditional
independence assumption \eqref{ci}, which relates model \eqref{530} to
model \eqref{520} through equation \eqref{535}. Equation~\eqref{535}
does not hold when assumption \eqref{ci} is violated. However, under
alternative assumptions, we have
%
%
\begin{equation}
\label{610} \bmalpha=\gamma\bigl(\beta_1,\bmbeta_X'
\bigr)'
\end{equation}
for a scalar $\gamma$. The key assumptions for \eqref{610} include
assumption \eqref{cire} and a \mbox{GLM-}like structure analogous to model
\eqref{520}:
%
%
\begin{equation}
\label{620} \pr\bigl(B=1|\bmX,U;\bmalpha^*\bigr)=\psi\bigl(
\alpha_1^*+\bolds{\beta}_X^{*\prime}\bmX+\alpha_U^*U\bigr),
\end{equation}
where $\bmalpha^*=({\alpha_1^*},\bmalpha_X^{*\prime},{\alpha_U^*})'$. In
Section~A of the supplemental article [\citet{z14}], we state
additional assumptions that lead to \eqref{610} and give an expression
for $\gamma$. Since $\bmbeta$ is identifiable and estimable using the
techniques described earlier, $\bmalpha$ can be estimated as soon as
$\gamma$ is known or estimated. Unfortunately, $\gamma$ is
unidentifiable from the observed data. For the probit and logit links,
we show in Section~A of the supplemental article [\citet{z14}]
that $\gamma$ can take any value greater than $2^{-1/2}\approx0.71$.
Thus, when assumption \eqref{ci} is in doubt, we can perform a
sensitivity analysis based on specified values of $\gamma\in
(2^{-1/2},\infty)$, with $\gamma=1$ corresponding to conditional independence.

\subsection{Indirect estimation of \texorpdfstring{$\pi_X(\mathbf x)$}{piX(x)} based on an outcome model}\label{indir}

An outcome model is a parametric model, say, $F_{Y|T,X}(y|t,\mathbf x
;\bmtheta)$, for the conditional distribution of $Y$ given $T$ and
$\bmX
$, specified up to a finite-dimensional parameter $\bmtheta$. Let
$f_{Y|T,X}(y|t,\mathbf x;\bmtheta)$ denote the associated probability
density or mass function. A typical outcome model would be a GLM with
the following mean structure:
%
%
\begin{equation}
\label{glm} \epn(Y|T,\bmX;\bmtheta)=\psi\bigl\{\theta_1+
\theta_TT+\bmtheta_X'\bmX+
\bmtheta_{TX}'(T\bmX)\bigr\},
\end{equation}
where $\bmtheta=(\theta_1,\theta_T,\bmtheta_X',\bmtheta_{TX}')'$
and $\psi$ is an inverse link function. The parameter $\bmtheta$ can be
estimated by maximizing the likelihood $\prod
_{i=1}^nf_{Y|T,X}(Y_i|T_i,\bmX_i;\bmtheta)$, and\vspace*{1pt} the resulting maximum
likelihood estimate (MLE) will be denoted by $\widehat\bmtheta$.
Because of randomization, the outcome model $F_{Y|T,X}(y|t,\mathbf x;\bmtheta
)$ implies that
\[
F_{t|X}(y|\mathbf x)=\pr\bigl\{Y(t)\le y|\bmX=\mathbf x\bigr
\}=F_{Y|T,X}(y|t,
\mathbf x;\bmtheta)\qquad(t=0,1),
\]
which can be estimated by substituting $\widehat\bmtheta$ for
$\bmtheta$.

Under the conditional independence assumption \eqref{ci}, the joint
distribution
\[
F_{\cdot|X}(y_0,y_1|\mathbf x)=\pr\bigl\{Y(0)\le
y_0,Y(1)\le y_1|\bmX=\mathbf x\bigr\}
\]
is identified as
\[
F_{0|X}(y_0|\mathbf x)F_{1|X}(y_1|
\mathbf x)=F_{Y|T,X}(y_0|0,\mathbf x;\bmtheta)F_{Y|T,X}(y_1|1,
\mathbf x;\bmtheta)
\]
and estimated by replacing $\bmtheta$ with $\widehat\bmtheta$. Write
\[
\widehat F_{\cdot|X}^{\mathrm{CI}}(y_0,y_1|\mathbf x)=F_{Y|T,X}(y_0|0,\mathbf x
;\widehat\bmtheta)F_{Y|T,X}(y_1|1,\mathbf x;\widehat\bmtheta),
\]
where the superscript CI stands for conditional independence. The
corresponding estimate of $\pi_X(\mathbf x)$ is then given by
\[
\int I\bigl\{(y_0,y_1)\in\mathcal B\bigr\}\widehat
F_{\cdot|X}^{\mathrm{CI}}(dy_0,dy_1|\mathbf x)=:
\widehat F_{\cdot|X}^{\mathrm{CI}}(\mathcal B|\mathbf x).
\]

When assumption \eqref{ci} is in doubt, we can perform a sensitivity
analysis based on assumption \eqref{cire}, which implies that
%
%
\begin{eqnarray}\label{indir10}
F_{\cdot|X}(y_0,y_1|\mathbf x) &=& \int F_{\cdot|X,U}(y_0,y_1|\mathbf x,u)F_{U}(du)
\nonumber\\[-8pt]\\[-8pt]\nonumber
&=& \int F_{0|X,U}(y_0|\mathbf x,u)F_{1|X,U}(y_1|
\mathbf x,u)F_{U}(du),
\end{eqnarray}
where we generalize the previous notation in an obvious way (with $U$
as an additional conditioning variable). This suggests that we specify
a model, say, $F_{Y|T,X,U}(y|t,\mathbf x,u;\bmtheta^*)$, for the
conditional
distribution of $Y$ given $(T,\bmX,U)$, with a finite-dimensional
parameter $\bmtheta^*$. Analogous to the GLM \eqref{glm}, we work with
a generalized linear mixed model (GLMM) with
%
%
\begin{equation}
\label{glmm} \epn\bigl(Y|T,\bmX,U;\bmtheta^*\bigr)=\psi\bigl\{
\theta_1^*+\theta_T^*T+\bmtheta_X^{*\prime}
\bmX+\bmtheta_{TX}^{*\prime}(T\bmX)+{\theta_U^*}U
\bigr\},
\end{equation}
where $\bmtheta^*=(\theta_1^*,\theta_T^*,\bmtheta_X^{*\prime},\bmtheta _{TX}^{*\prime},{\theta_U^*})'$. The GLMM is not completely identified
without additional information, and we propose a sensitivity analysis
based on specified values of ${\theta_U^*}$ (or, rather, $|\theta
_U^*|$), which is described in Section~B of the supplemental article
[\citet{z14}].

It is worth mentioning that the random effect $U$ has a different
interpretation here than in Section~\ref{dir}. In model \eqref{glmm},
$U$ represents an unobserved prognostic factor which affects both
potential outcomes in the same direction; a change in $U$ may or may
not have much effect on the treatment benefit $B$, depending on the
precise definition of $B$ and model \eqref{glmm}. Although one could
incorporate a random treatment effect into model \eqref{glmm}, the
resulting method will likely become very complicated. In model \eqref
{620}, $U$ acts like an effect modifier in that a change in $U$ leads
directly to a change in the probability of a desirable treatment
benefit. (Here we use the term effect modifier in a heuristic sense
which may or may not agree with an interaction-based definition.) Thus,
aside from modeling assumptions, the direct and indirect approaches
also differ in how they deal with departures from assumption \eqref
{ci}. The indirect approach is designed to address violations of
assumption~\eqref{ci} due to an unmeasured prognostic factor, whereas
the direct approach is more appropriate for violations of assumption
\eqref{ci} due to an unmeasured effect modifier.

\subsection{Estimation of \texorpdfstring{$\pi_Z(z)$}{piZ(z)}}\label{bet}

Equations \eqref{30} and \eqref{40} suggest that evaluation of a
predictive biomarker $Z$ can be based on $\pi_Z(z)=\pr(B=1|Z=z)$ and
the marginal distribution of $Z$. Because the latter is straightforward
to estimate, this section is focused on estimation of $\pi_Z(z)$ with a
given estimate of $\pi_X(\mathbf x)$, say, $\widehat\pi
_X(\mathbf x)$, which may
be obtained using any one of the proposed methods. For a binary marker,
equation \eqref{50} suggests that $\pi_Z(z)$ can be estimated by
$\sum
_{i=1}^nI\{Z_i=z\}\widehat\pi_X(z,\bmW_i)/\sum_{i=1}^nI\{Z_i=z\}$. We
therefore assume that $Z$ is continuous in the rest of this section.

We propose to estimate $\pi_Z(z)$ by substituting an estimate of
$F_{W|Z}$ into equation~\eqref{50}. One could specify a parametric
model for $F_{W|Z}$, however, this can be difficult because the
dimension of $\bmW$ can be rather large (5 in our example). We
therefore exploit the fact that $Z$ is only one-dimensional and employ
nonparametric regression techniques in estimating $F_{W|Z}$. Let $\xi
\dvtx \mathbb R\to[0,\infty)$ be a kernel function and $\lambda>0$ a
bandwidth parameter. Then we can estimate $F_{W|Z}(\mathbf w|z)$ by
\[
\frac{\sum_{i=1}^n\xi\{(Z_i-z)/\lambda\}I(\bmW_i\le\mathbf w)}{\sum
_{i=1}^n\xi\{(Z_i-z)/\lambda\}}.
\]
This, together with $\widehat\pi_X(\mathbf x)$, can be
substituted into
equation \eqref{50} to estimate $\pi_Z(z)$ as
\[
\widehat\pi_Z(z)=\frac{\sum_{i=1}^n\xi\{(Z_i-z)/\lambda\}\widehat
\pi
_X(z,\bmW_i)}{\sum_{i=1}^n\xi\{(Z_i-z)/\lambda\}}.
\]

An important question here is how to choose the bandwidth $\lambda$,
for which we propose a cross-validation approach. The estimate
$\widehat
\pi_Z(z)$ can be regarded as a nonparametric regression of $\widehat
\pi
_X(Z,\bmW)$ on $Z$, and its\vspace*{1pt} predictive accuracy can be assessed by
comparing the ``response'' $\widetilde B_i=\widehat\pi_X(\bmX_i)$ with
the estimate $\widehat B_i=\widehat\pi_Z(Z_i)$. We\vspace*{1pt} propose to partition
the sample into a training set and a validation set and choose a value
of $\lambda$ that minimizes the average of $(\widetilde B_i-\widehat
B_i)^2$ in the validation set with $\widehat\pi_Z(z)$ estimated from
the training set using bandwith $\lambda$.

\subsection{Estimation of TPR, FPR and ROC}\label{poi}

Given $\widehat\pi_Z(z)$ from Section~\ref{bet}, the parameters of
interest can be estimated using equations \eqref{30} and \eqref{40}.
For a binary marker, this leads to
\begin{eqnarray*}
\widehat{\mathrm{TPR}}&=&\frac{\widehat\tau\widehat\pi
_Z(1)}{\widehat\tau
\widehat\pi_Z(1)+(1-\widehat\tau)\widehat\pi_Z(0)},
\\
\widehat{\mathrm{FPR}}&=&\frac{\widehat\tau\{1-\widehat\pi_Z(1)\}
}{\widehat\tau\{1-\widehat\pi_Z(1)\}+(1-\widehat\tau)\{1-\widehat
\pi
_Z(0)\}},
\end{eqnarray*}
where $\widehat\tau=n^{-1}\sum_{i=1}^nZ_i$. For a continuous marker,
we have
\[
\widehat{\mathrm{ROC}}(s)=1-\widehat F_{Z|B=1}\bigl\{\widehat
F_{Z|B=0}^{-1}(1-s)\bigr\},
\]
where
\begin{eqnarray*}
\widehat F_{Z|B=1}(z_0)&=&\sum_{i=1}^nI(Z_i
\le z_0)\widehat\pi_Z(Z_i) \Big/\sum
_{i=1}^n\widehat\pi_Z(Z_i),
\\
\widehat F_{Z|B=0}(z_0)&=&\sum_{i=1}^nI(Z_i
\le z_0)\bigl\{1-\widehat\pi_Z(Z_i)\bigr\}
\Big/\sum_{i=1}^n\bigl\{1-\widehat
\pi_Z(Z_i)\bigr\}.
\end{eqnarray*}

An asymptotic analysis of these estimates can be rather tedious,
especially because $\widehat\pi_Z(z)$ involves smoothing and
cross-validation. For inference, we recommend the use of bootstrap
confidence intervals. To account for all variability in the estimates,
the entire estimation procedure, including bandwidth selection based on
cross-validation, should be repeated for each bootstrap sample.

\section{Application to the THRIVE study}\label{app}

We now apply the methods of Section~\ref{meth} to the THRIVE study
introduced in Section~\ref{intro}, a randomized, double-blind,
double-dummy, noninferiority trial at 98\vadjust{\goodbreak} hospitals or medical centers
in 21 countries [\citet{c11}]. The THRIVE study compared
rilpivirine with efavirenz for treating HIV-1 in treatment-naive
adults, in the presence of common background nucleoside or nucleotide
reverse transcriptase inhibitors (N[t]RTIs). The study enrolled 680
adult patients who were naive to antiretroviral therapy, with a
screening viral load of at least 5000 copies per ml and viral
sensitivity to N(t)RTIs. The patients were randomized in a $1\dvtx 1$ ratio
to receive oral rilpivirine 25 mg once daily or efavirenz 600 mg once
daily, in addition to an investigator-selected regimen of background N(t)RTIs.

The outcome of interest to us is virologic response (viral load below
50 copies${}/{}$ml) at week 48 of treatment, with patient discontinuation
(about 5\%) counted as failure. The observed virologic response rates
are 86\% and 82\% in the rilpivirine and efavirenz groups,
respectively, and the difference between the two groups (3.5\%; 95\%
CI: $-$1.7--8.8\%) meets a prespecified noninferiority criterion based
on a 12\% margin ($p<0.0001$). Thus, rilpivirine appears comparable to,
if not better than, efavirenz in terms of population-level efficacy.
However, Figure~\ref{fig1} suggests that individual patients respond differently
to the two therapies and that baseline viral load and CD4 cell count
could be used as predictive biomarkers. As indicated earlier, we will
for safety reasons define individual-level treatment benefit as $B=I\{
Y(0)\le Y(1)\}$, where $Y(0)$ and $Y(1)$ denote potential outcomes for
efavirenz and rilpivirine, respectively.

Baseline viral load and CD4 cell count are both log-transformed before
entering the benefit and outcome models as covariates. In addition to
these biomarkers, the covariate vector $\bmX$ also includes gender,
race (black, white or other), age and body mass index at baseline. For
the direct approach of Section~\ref{dir}, the benefit model is a
logistic regression model given by \eqref{520}, with the aforementioned
covariates as linear terms (and no interactions), and the helper model
is given by \eqref{530} with the same link. For the indirect approach
of Section~\ref{indir}, the outcome model is a logistic regression
model similar to \eqref{glm} except that interactions of $\bmX$ with
$T$ are limited to the two biomarkers. The selection of covariates and
interactions in these models is based on subject matter knowledge and
not on statistical tests. Estimation of model \eqref{530} is based on
the 1\% of pairs (of control and experimental patients) that are most
similar in terms of $\llVert\bmX_i-\bmX_j\rrVert$, as suggested by
\citet{d03}. (Other proportions, from 0.001 to 1, have been
attempted without yielding a material difference.) Under the indirect
approach, the simplified method in Web Appendix B is used to estimate
$\bmtheta^*$ for a given $\theta_U^*$. In any case, estimates of $\pi
_X(\mathbf x)$ are converted into estimates of $\pi_Z(z)$ using
the kernel
smoothing method of Section~\ref{bet}, with a Gaussian kernel and a
cross-validated bandwidth. The cross-validation is based on a $1\dvtx 1$
random partition of the sample into a training set and a validation
set, and the bandwidth is chosen among $\{2^l\sd(Z)\dvtx l=-5,\ldots,5\}$
using the minimum mean squared error criterion. The formulas of
Section~\ref{poi} are used to obtain estimates of ROC curves, and the
trapezoidal rule is then employed in calculating the associated AUCs.
The above procedure is performed for both biomarkers on the original
sample as well as 200 bootstrap samples. Pointwise 90\% confidence
intervals for ROC curves are obtained using a simple bootstrap
percentile method, and bootstrap standard errors are used for inference
on AUCs (and AUC differences between the two biomarkers).

%
%
\begin{figure}

\includegraphics{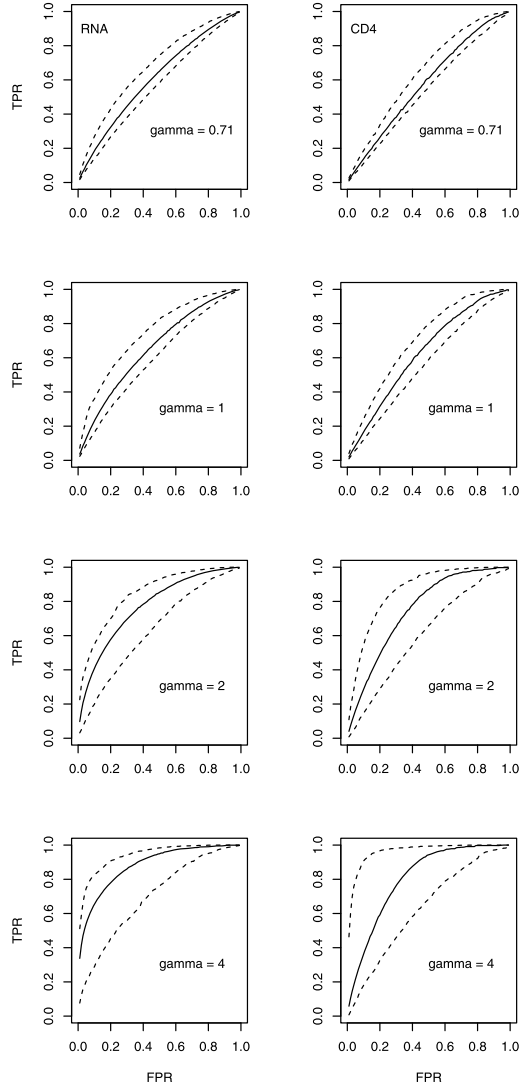}

\caption{ROC analysis of the THRIVE data using the direct approach:
estimated ROC curves (solid lines) and 90\% pointwise confidence
intervals (dashed lines) for baseline viral load (RNA, left) and CD4
cell count (right), with different values of $\gamma$.}\label{fig2}
\end{figure}

Figure~\ref{fig2} gives a side-by-side comparison of ROC curves for the two
biomarkers, estimated using the direct approach of Section~\ref{dir}
with $\gamma=0.71,1,2,4$. The value $\gamma=1$ corresponds to
assumption \eqref{ci} of conditional independence, while the value 0.71
is a theoretical lower bound. The associated AUC estimates and standard
errors are shown in the upper half of Table~\ref{tb1}. Both Figure~\ref{fig2} and the
relevant portion of Table~\ref{tb1} show that the two biomarkers are useful as
predictive biomarkers, with ROC curves above the diagonal and AUCs
greater than 0.5 after accounting for sampling variability. The
performance of each biomarker does appear to depend heavily on the
value of $\gamma$; the AUC estimate increases dramatically with
increasing $\gamma$. This pattern is confirmed by additional analyses
based on other values of $\gamma$ (results not shown). Given the
remarks at the end of Section~\ref{indir}, these results suggest that
evaluation of a predictive biomarker can be rather sensitive to an
unmeasured effect modifier. On the other hand, for each value of
$\gamma
$, the AUC estimate for baseline viral load is higher than that for
baseline CD4 cell count, although the difference is not statistically
significant. These results suggest that comparison of predictive
biomarkers may be less sensitive to the choice of $\gamma$, even though
this particular data set does not provide strong evidence that baseline
viral load is better than baseline CD4 cell count as a predictive
biomarker. Whether we are evaluating a single marker or comparing two
markers, there is an obvious relationship between increasing $\gamma$
and greater variability in the estimates, as indicated by widening
confidence intervals in Figure~\ref{fig2} and increasing standard errors in Table~\ref{tb1}.

%
\begin{table}
\tabcolsep=3pt
\caption{AUC analysis of the THRIVE data: point estimates and bootstrap
standard errors for the AUCs of baseline viral load (RNA) and CD4 cell
count as well as their difference (RNA--CD4),
obtained using the direct approach of Section~\protect\ref{dir} and the
indirect approach of Section~\protect\ref{indir} with sensitivity
parameters $\gamma$ and $\theta_U^*$, respectively}\label{tb1}
\begin{tabular*}{\tablewidth}{@{\extracolsep{\fill}}@{}lcccccc@{}}
\hline
\multirow{3}{41pt}{\textbf{Sensitivity parameter}} & \multicolumn{3}{c}{\textbf{Point estimate}} & \multicolumn{3}{c@{}}{\textbf{Standard error}}\\[-6pt]
& \multicolumn{3}{c}{\hrulefill} & \multicolumn{3}{c@{}}{\hrulefill}\\
& \multicolumn{1}{c}{\textbf{RNA}} & \multicolumn{1}{c}{\textbf{CD4}} &\multicolumn{1}{c}{\textbf{Diff.}}& \multicolumn{1}{c}{\textbf{RNA}} & \multicolumn{1}{c}{\textbf{CD4}} & \multicolumn{1}{c@{}}{\textbf{Diff.}}\\
\hline
$\gamma$ &\multicolumn{6}{c@{}}{\textit{Direct approach}}\\
0.71&0.61&0.58&0.03&0.04&0.03&0.03\\
1&0.65&0.63&0.03&0.05&0.04&0.05\\
2&0.77&0.75&0.02&0.06&0.08&0.08\\
4&0.88&0.80&0.08&0.07&0.10&0.13
\\[3pt]
$\theta_U^*$ & \multicolumn{6}{c@{}}{\textit{Indirect approach}}\\
0&0.64&0.59&0.05&0.04&0.04&0.05\\
1.8&0.65&0.60&0.06&0.05&0.05&0.05\\
4&0.68&0.64&0.04&0.07&0.07&0.08\\
8&0.72&0.70&0.02&0.09&0.09&0.11\\
\hline
\end{tabular*}
\end{table}

%
%
\begin{figure}

\includegraphics{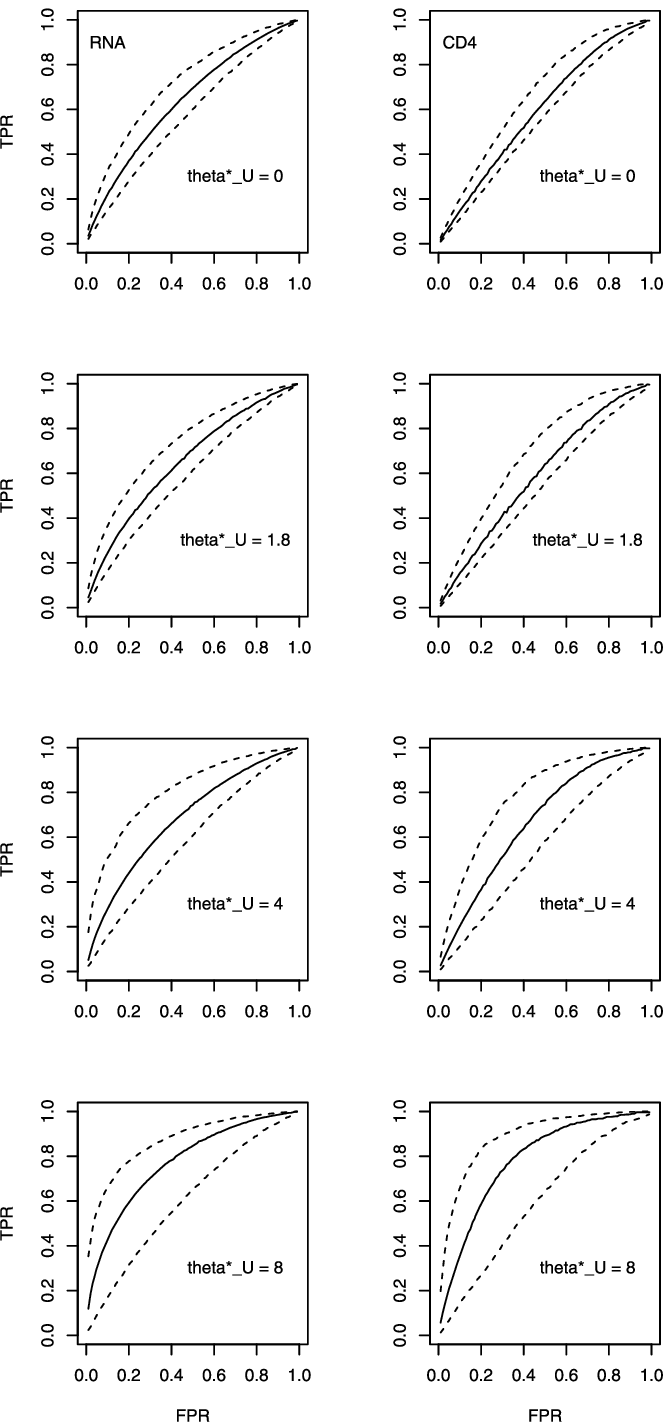}

\caption{ROC analysis of the THRIVE data using the indirect approach:
estimated ROC curves (solid lines) and 90\% pointwise confidence
intervals (dashed lines) for baseline viral load (RNA, left) and CD4
cell count (right), with different values of $\theta_U^*$.}\label{fig3}
\end{figure}

Figure~\ref{fig3} gives another comparison of the two biomarkers based on ROC
curves estimated using the indirect approach of Section~\ref{indir}
with $\theta_U^*=0,1.8,\break 4,8$. Here, the value $\theta_U^*=0$ corresponds
to conditional independence, and the value 1.8 is a lower confidence
bound (to be discussed later). The associated AUC estimates and
standard errors are shown in the lower half of Table~\ref{tb1}. These results
are consistent with those from the direct approach in suggesting that
both biomarkers are useful as predictive biomarkers. In particular, the
results for conditional independence ($\theta_U^*=0$) are fairly
consistent with their counterparts under the direct approach (with
$\gamma=1$). Like Figure~\ref{fig2}, Figure~\ref{fig3} also shows an increasing trend for
the predictive performance of each biomarker as a function of the
sensitivity parameter~$\theta_U^*$. The trend is confirmed for
additional values of $\theta_U^*$, although the results are not shown.
Thus, considering the remarks at the end of Section~\ref{indir}, it
appears that evaluation of a predictive biomarker can also be sensitive
to an unmeasured prognostic factor. The results from the indirect
approach reinforce the previous observation that baseline viral load
appears to perform better than baseline CD4 cell count, although the
differences here also fail to reach statistical significance. As is the
case with the direct approach, increasing $\theta_U^*$ tends to produce
greater variability in the estimates.

Although the uncertainty in comparing the two biomarkers could
potentially be reduced by an increased sample size, some uncertainty
will likely remain in evaluating each individual marker, given the
apparent dependence on sensitivity parameters. Nonetheless, the
increasing trend observed under both approaches suggests that a lower
bound for predictive performance may be available under each approach.
For the direct approach, the lower bound is given by $\gamma=0.71$, as
noted in Section~\ref{dir}. For the indirect approach, a lower bound
for $\theta_U^*$ (and hence for the performance of each biomarker) may
be available from longitudinal data (see Web Appendix B). For the
THRIVE study, the lower bound for $\theta_U^*$ is estimated as 2.5
(95\% CI: 1.8--3.7, based on 1000 bootstrap samples) from a GLMM
analysis of repeated measurements at 24, 32, 40 and 48 weeks. Although
earlier measurements (baseline through 20 weeks) are also available, we
restrict our analysis to the later measurements in order to reduce
misspecification bias; see \citeauthor{z13} [(\citeyear{z13}), Section~4] for a detailed
explanation of this strategy. This GLMM analysis suggests that, under
the additional assumptions given in Web Appendix B, the value $\theta
_U^*=1.8$ represents the worst case scenario for the indirect approach.
The corresponding ROC and AUC estimates are better than those for
$\gamma=0.71$ under the direct approach and thus more informative as
lower bounds.

Our ROC analyses under both (direct and indirect) approaches also
illustrate that measures of predictive accuracy need not correlate with
interactions. Although baseline CD4 cell count exhibits a more dramatic
interaction in Figure~\ref{fig1}, there is no indication (in the same data set)
that it outperforms baseline viral load as a predictive biomarker.

\section{Discussion}\label{disc}

In this article we have proposed new methods for evaluating predictive
biomarkers in the potential outcome framework of \citet{h12}.
Instead of monotonicity, our starting point is conditional independence
of potential outcomes given observed covariates. Possible departures
from the latter assumption can be addressed by incorporating a random
effect that accounts for any residual dependence between potential
outcomes. Because the random effect models are not completely
identifiable, we propose a sensitivity analysis approach based on
quantities related to the random effect. Our analysis of the THRIVE
data reveals a great deal of sensitivity for the performance of each
individual biomarker and much less sensitivity for the comparison of
the two markers. Despite the uncertainty about individual biomarkers,
the lower bounds on their predictive performance, available under both
(direct and indirect) approaches, show that they are both useful as
predictive biomarkers. For comparing the two markers, our analysis does
not show much sensitivity and does not indicate a significant
difference in the predictive performance of the two markers.

There does appear to be a lot of sensitivity (particularly in the upper
portion) in quantifying the performance of an individual marker. Such
sensitivity introduces additional uncertainty into the overall
conclusion of the analysis, which is certainly undesirable. We see this
as a reminder of the inherent limitation of the observed data for
answering certain questions. As in many other statistical problems
(e.g., missing data, censoring, confounding), the parameters of
interest to us are not identifiable from the observed data without
making untestable assumptions. When such assumptions are in doubt, the
parameters are partially identified and the associated inference cannot
be as sharp as that for point-identified parameters [\citet{m03}].
It may be disappointing to see that different assumptions (about the
sensitivity parameter) can lead to a wide range for the parameter of
interest. On the other hand, one could argue that the sensitivity
analysis serves its purpose well by revealing the limitation of the
available data and information.

In theory, point identification is achieved under the conditional
independence assumption (6), which requires that the observed
covariates be sufficient for explaining the dependence between the
potential outcomes. Although we may never be certain that assumption
(6) holds, it seems reasonable to believe that the assumption will get
close to being true with a growing set of relevant covariates obtained
from increasing knowledge of the disease. The more we know about the
disease, the more information we have about relevant covariates, the
more confidence we should be able to place in assumption (6). In
practice, it may be difficult to determine when we have sufficient
information to rely on assumption (6) and when we have to perform a
sensitivity analysis. One possible solution would be a Bayesian
approach with an informative prior on the sensitivity parameter which
quantifies our uncertainty about assumption (6). As we become more
confident about assumption (6), the prior will become more concentrated
at or near that assumption.

Each of the direct and indirect approaches has unique advantages. The
direct approach is able to accommodate complex definitions of treatment
benefit involving several outcome variables of arbitrary types, as long
as they are all observed. The indirect approach is best suited for a
single outcome of primary interest. Although the outcome model can
include multiple outcomes in principle, their dependence structure can
be difficult to specify and estimate. Even for a single outcome, the
indirect approach requires a greater amount of modeling (in the sense
that an outcome model implies a benefit model) and is therefore more
prone to misspecification bias. On the other hand, the indirect
approach is able to use all the information in the observed outcome
data and therefore may have an efficiency advantage. Finally, for
sensitivity analysis, the direct approach is more appropriate for an
unmeasured effect modifier, and the indirect approach for an unmeasured
prognostic factor. In practice, we recommend that both approaches be
used in a complementary fashion, as in our analysis of the THRIVE data.

A reviewer has pointed out that some elements of personal judgment may
be involved in choosing among different treatments. While this is not a
major issue in the THRIVE data, where the definition of a treatment
benefit is quite objective, it can certainly become a major issue in
other therapeutic areas such as weight loss. For example, some patients
may be willing to accept the extra risks of a surgical procedure
(relative to a nonsurgical one) for an additional loss of 20 pounds,
while others may not. To incorporate such personal judgment into the
proposed approach, we could allow $\delta$ to vary across patients, and
we would need to be able to measure $\delta$ for individual patients or
at least be able to predict $\delta$ using measurable individual
characteristics. In the latter case, the methodology will need to be
modified to incorporate a prediction model for $\delta$ and the
associated variability. It will be of interest to explore that
possibility in the context of a suitable application.

Because of the complexity of the proposed methodology, a sample size
formula is not yet available; however, for a given application, one
could perform a simulation study to gauge the adequacy (in terms of
power and precision) of a proposed sample size. Such an assessment
should obviously consider the objective of the analysis (e.g.,
evaluating a single biomarker versus comparing two or more biomarkers).
In addition, the nonparametric regression in Section~\ref{bet} may
imply a higher requirement on the sample size than do the other parts
of the methodology, which are based on parametric regression
techniques. A sample size that is inadequate for one-dimensional
nonparametric regression may compromise the performance of the methods.

\section*{Acknowledgment}
We thank the Editor (Dr. Karen Kafadar), the Associate Editor and three
anonymous reviewers for their insightful and constructive comments,
which have greatly improved the manuscript. The views expressed in this
article do not represent the official position of the U.S.~Food and
Drug Administration.

\begin{supplement}[id=suppA]
\sname{Supplement}
\stitle{Technical details}
\slink[doi]{10.1214/14-AOAS773SUPP} 
\sdatatype{.pdf}
\sfilename{AOAS773\_supp.pdf}
\sdescription{We provide technical details concerning the
sensitivity analyses in Sections~\ref{dir} and~\ref{indir}.}
\end{supplement}

%

\printaddresses
\end{document}